\documentstyle[pra,aps,epsfig]{revtex}
\begin{document}
\draft \twocolumn[\hsize\textwidth\columnwidth\hsize\csname @twocolumnfalse\endcsname
\title{Quantum phase gate with a selective interaction}
\author{E. Solano$^{*}$$^{1,2}$, M. Fran\c ca Santos$^1$, and
P. Milman$^{**}$$^{1}$}

\address{$^1$Instituto de F\'{\i}sica, Universidade Federal do Rio de Janeiro,
Caixa Postal 68528, 21945-970 Rio de Janeiro, RJ, Brazil \\
$^{2}$Secci\'{o}n F\'{\i}sica, Departamento de Ciencias,
Pontificia Universidad Cat\'{o}lica del Per\'{u}, Apartado 1761,
Lima, Peru}
\maketitle

\begin{abstract}
We present a proposal for implementing quantum phase gates using
selective interactions. We analize selectivity and the possibility
to implement these gates in two particular systems, namely, trapped ions
and Cavity QED.
\end{abstract}
 \vskip1pc]

Quantum computers would perform
certain tasks, such as factoring a number and searching a data in
a array, faster than a classical computer~\cite{bennettcomp,ekexxx95}. The
core of quantum computing are the quantum logic gates. In fact, it
is known that any quantum computation can be reduced to a sequence
of universal two qubit logic gates and one qubit local
operations~\cite{divxxx95}. Since the original formulation  of
quantum computation, a number of experimental systems have been
proposed as candidates for the practical implementation of these
quantum gates. We may mention trapped ions \cite{ciraczoller1},
cavity QED \cite{realizable,tombesi}, liquid state nuclear
magnetic resonance (NMR) \cite{nmr}, quantum dots \cite{dots},
optical lattices \cite{caves}, among others. These systems have
shown to be good enough not only for testing quantum logic gates
but also for some basic quantum operations with a few qubits. For
example, the realization of quantum logic gates in trapped ions
\cite{wineland}, cavity QED \cite{qpgparis,qpgkimble} and NMR
\cite{nmrexp} have already been possible. The aim of the mentioned
experiments was essentially the practical realization in a
bipartite system of two equivalent kinds of universal two qubit
logic gates: a quantum controlled not gate (CNOT) and a quantum
phase gate (QPG)~\cite{preskill}. The CNOT gate and the QPG differ
from each other only by local operations (single qubit rotations).
In a given bipartite system, consisting of the so called control
and target qubits, a CNOT operation changes the target qubit state
only if the control qubit state is in a specific state.
Explicitly, a CNOT operation acting on the initial arbitrary state
\begin{eqnarray}\label{cnot}
|\Psi \rangle = \alpha |\downarrow,\downarrow \rangle + \beta
|\uparrow,\downarrow \rangle+\gamma |\downarrow,\uparrow \rangle
+\delta |\uparrow,\uparrow \rangle
\end{eqnarray}
produces
\begin{eqnarray}\label{cnot}
|\Psi^{\prime} \rangle = \alpha |\downarrow,\downarrow \rangle +
\beta |\uparrow,\uparrow \rangle+\gamma |\downarrow,\uparrow
\rangle + \delta |\uparrow,\downarrow \rangle ,
\end{eqnarray}
where the first label in the kets refers to the control qubit and
the second one refers to the target qubit. On the other hand, a
QPG acting on the same initial state produces
\begin{eqnarray}\label{qpg}
&&|\Psi^{\prime \prime} \rangle =  - \alpha |\downarrow,\downarrow
\rangle + \beta |\uparrow,\downarrow \rangle+\gamma
|\downarrow,\uparrow \rangle + \delta |\uparrow,\uparrow \rangle.
\end{eqnarray}
It can be easily seen that a QPG can act as a CNOT gate if we
rotate the target bit before and after operating the gate,
according to the following steps

i) A single qubit rotation in the target qubit
\begin{eqnarray}
|\downarrow \rangle \rightarrow | - \rangle = \lbrack |\downarrow
\rangle - |\uparrow \rangle \rbrack / \sqrt{2} \nonumber \\
|\uparrow \rangle \rightarrow | + \rangle = \lbrack |\downarrow
\rangle + |\uparrow \rangle \rbrack / \sqrt{2} ,
\end{eqnarray}
also known as Hadamard gate.

ii) A QPG.

iii) Another Hadamard rotation in the target qubit.

\noindent Because of their fundamental interest, in quantum logic
tests as well as in the search of scalable quantum computing, it
is always useful to find ways of implementing CNOT and QPG gates
in the laboratory. This fact motivated a number of proposals for
realizing such gates in different quantum systems. The main
problem these systems face, when scalability is the goal, is
decoherence. For the quantum computing schemes to work it is
essential to keep coherence of the qubits themselves and among
them. However, when the dimension of the system and the number of
operations increase decoherence effects can become more and more
important. Therefore, simplifying the operations on the qubits
\cite{sorensen} as well as making faster logic gates \cite{daniel}
are the main purpose of many of these works.

In this paper, we will be concerned with turning the computational
process simpler: we propose the use of a selective interaction
that would realize a quantum phase gate with a single pulse, i.e.,
without changing experimental parameters during the process. We
will discuss our method in a trapped ions system and in the domain
of CQED.

First, we will consider an array of N ions of mass $m$ in a Paul trap.
The ions will be treated as two level systems that interact with
each other through a coulombian force, so that collective
vibrational modes can be conveniently
introduced~\cite{james}. The manipulation of both electronic states and
vibrational collective modes can be done by means of laser beams
tuned to appropriate frequencies.

Our task is then to produce an interaction that performs the
transformation described in Eq.~(\ref{qpg}). This interaction must act
effectively on two chosen ions ($j$ and $k$) in the array, ion $j$
being the control qubit and ion $k$ the target qubit. To achieve
this, we address ion $j$ with a Raman laser pair described by the
electric field $\vec{E}_{I}= \vec{E}_{oI}e^{i (qz - \omega
_{I}t)}$ and ion $k$ with another Raman laser pair described by
the electric field $\vec{E}_{II}= \vec{E}_{oII}e^{i (qz - \omega
_{II}t)}$ (see Fig.~1). The Raman systems have different effective
frequencies, $\omega_I$ and $\omega_{II}$, that are quasiresonant
to an electronic transition associated with the angular frequency
$\omega_{o}$. The Hamiltonian corresponding to this situation is
\begin{eqnarray}
\label{hamion} \hat H & = &\hat H_{o} + \hat H_{{\rm int}} ,
\end{eqnarray}
with
\begin{eqnarray}
\hat H_{o} = && \hbar \omega _{o}\left(\hat S_{+j}\hat S_{-j} +
\hat S_{+k} \hat S_{-k}\right) \nonumber \\  && + \hbar \nu \hat
a^{\dagger } \hat a + \sum_{n} \hbar \nu _{n} \hat b_n^{\dagger }
\hat b_n
\end{eqnarray}
and
\begin{eqnarray}
\label{hint}
\hat H_{{\rm int}} = && \hbar \Omega \lbrace ( \hat
S_{+j} +\hat S_{-j} ) ( e^{i (q \hat z_{j} - \omega _{I} t ) } +
e^{-i (q \hat z_{j} - \omega _{I} t ) } )  \nonumber \\  && + (
\hat S_{+k} +\hat S_{-k} ) ( e^{i (q \hat z_{k} - \omega _{II} t )
} + e^{-i (q \hat z_{k} - \omega _{II} t ) } ) \rbrace
\end{eqnarray}
Here, $\widehat{S}_{+i}=|\uparrow _{i}\rangle \langle \downarrow
_{i}|$, $\widehat{S}_{-i}=|\downarrow _{i}\rangle \langle \uparrow
_{i}|$ and the state $|\uparrow_i \rangle$ ($|\downarrow_i
\rangle$) corresponds to the $i$-th ion in the excited
(fundamental) state. $\hat{a}^{\dagger}$ ($\hat{a}$) is the
creation (annihilation) operator associated with the harmonic
oscillation, with frequency $\nu$, of the CM mode.
$\hat{b}^{\dagger}_{n}$ ($\hat{b}_{n}$) is the creation
(annihilation) operator associated with the harmonic oscillation,
with frequency $\nu_{n}$, of the other collective modes. All
frequencies $\nu_{n}$ are bigger than $\nu$~\cite{james}.
$\hat{z}_{j}$ and
$\hat{z}_{k}$ are the operators corresponding to the positions of
the ions $j$ and $k$, respectively, and can be rewritten as linear
combinations of the operators corresponding to the collective
coordinates. $\hat H_{o}$ is the free Hamiltonian that corresponds
to the internal energy of the two ions plus the energy of the CM
mode and of the other collective modes. $\hat H_{int}$ is the
interaction Hamiltonian describing the position dependent dipolar
interaction of ions $j$ and $k$ with the two Raman beams.

\vspace*{1cm}

\begin{figure}
\hspace*{1cm} \leavevmode \epsfxsize=5 cm \epsfbox{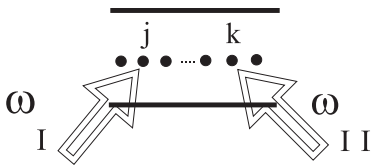}
\vspace*{1cm} \caption{Two Raman beams of frequencies $\omega_I$
and $\omega_{II}$ interacting dispersively with ions $j$ and $k$,
respectively.} \label{fig1}
\end{figure}

We choose the frequencies $\omega_{I}$ and $\omega_{II}$ to be
quasiresonant to the first upper CM sideband and to the carrier,
respectively. Specifically, $\omega_{I} = \omega_{o} + \nu -
\delta$ and $\omega_{I} = \omega_{o} + \delta$, where $\delta$ is
the detuning of each Raman beam with respect to the mentioned
resonant vibronic transitions, such that $\omega_{I} +\omega_{II}
= 2 \omega_{o} + \nu$. Similarly to what was done in
\cite{kikebell,nos}, the Hamiltonian in Eq.~(\ref{hint}) can be
expanded in terms of creation and anihilation operators of the
normal modes and rewritten in the interaction picture. Then,
following the standard procedure described in~\cite{rwaions}, we
can make the rotating wave approximation (RWA) and discard the
terms that oscillate with higher frequencies in the dispersive
limit $\Omega \ll \delta \ll \nu$. In this way, the following
effective time independent Hamiltonian
\begin{eqnarray}\label{hcm}
&&H_{{\rm eff}} = \hbar \Omega _{o}\hat{G}_{o}^{2}\{ i \eta
\hat{S}_{+j} \hat{S}_{+k} {\rm e}^{2 i \phi}[\hat{a}^{\dagger }
\hat{F}_{o} - \hat{F}_{o} \hat{a}^{\dagger} ]\hat{F}_{1}
\nonumber\\ && + \frac{1}{2}\hat{S}_{+j}\hat{S}_{-j}[\eta
^{2}\hat{a}^{\dagger}\hat{F}_{1}^{2}\hat{a}]-\frac{1}{2}
\hat{S}_{-j}\hat{S}_{+j}[\eta^2 \hat{F}_{1}  \hat{a}
\hat{a}^{\dagger} \hat{F}_{1} ] \nonumber \\ &&-\frac{1}{2}
\hat{S}_{+k}\hat{S}_{-k}[\hat{F}_{o}^{2}]+\frac{1}{2}
\hat{S}_{-k}\hat{S}_{+k}[\hat{F}_{o}^{2}]+{\rm H.c.}\} \,
\end{eqnarray}
can be derived, as was done in \cite{kikebell,nos}. Here,
$\Omega_o=\frac{\Omega^2}{\delta}$ and $\eta=q\sqrt{\hbar/2 N m
\nu}$ is the Lamb-Dicke parameter associated to the CM motion. The
functions
\begin{eqnarray}
\hat{F}_{k} &=&\sum_{n} f(n) |n\rangle \langle n| ,
\end{eqnarray}
with
\begin{eqnarray}
f(n) = e^{-\eta ^{2}/2}\frac{n!}{(n+k)!} L_{n}^{k}(\eta^{2}) ,
\end{eqnarray}
where $L_{n}^{k}(\eta^{2})$ are the generalized Laguerre
polynomials, can always be written in terms of the CM mode number
operator $\hat{n}$. $\hat{G}_{o}$ is a similar non-linear function
involving the number operators related to all other normal modes.
The exact form of function $\hat{G}_{o}$ is irrelevant in our case,
since we will
suppose in this paper that all collective modes are cooled down
to the ground state and that only the CM mode will be excited. In
this case $\hat{G}_{o}$ contributes to the effective Rabi frequency with only
a constant numerical factor of the order of $1$. The first term of
$H_{eff}$ and its Hermitian conjugate describe the common
excitation of both electronic states and the CM mode, similar to a
non linear anti-Jaynes-Cummings interaction. The other terms are
motional dependent self energy terms. The main difference of the
Hamiltonian of Eq.~(\ref{hcm}), when compared to the one described
in Ref.~\cite{nos}, is that now the subspace
$\{|\downarrow\uparrow \rangle,|\uparrow\downarrow\rangle \}$
remains untouched, a natural consequence of the ionic individual
addressing demanded in the present scheme.

Selectivity arises from the $\hat n$ dependence of the self energy
corrections in Eq.~(\ref{hcm}). The stark shift of the electronic
states of the ions depends explicitly on the number of phonons of
the CM mode through $\hat{F}_{0}^2$. We can adjust the laser beams
frequencies to compensate this shift for one particular subspace transition
tuning it to resonance. The new
frequencies depend strongly on the selected vibronic subspace we
want to excite. This will yield another selective interaction,
different from the one discussed in \cite{nos}, that only performs
resonant transitions inside the closed subspace
$\{|\downarrow\downarrow \rangle |n
\rangle,|\uparrow\uparrow\rangle |n+1 \rangle\}$ without producing
transitions inside the subspace $\{|\downarrow\uparrow \rangle,|\uparrow\downarrow\rangle \}$. For the sake of
simplicity, we only label the CM mode excitacions, since the other
vibronic states are cooled down to the vacuum state and, in this way,
are not affected by the interaction Hamiltonian.

We are now ready to show how a QPG can be implemented using the
interaction described in the Hamiltonian of Eq.~(\ref{hcm}). We
study the effect of the associated evolution operator $\hat U$
over some relevant states of the Hilbert space
\begin{eqnarray}
\label{evolucao} \hat U|\downarrow\downarrow,0
\rangle=&&\cos(\Omega_{eff} t) |\downarrow\downarrow,0 \rangle +
\sin(\Omega_{eff} t) |\uparrow\uparrow,1 \rangle \nonumber \\
U|\uparrow\downarrow,0 \rangle=&&|\uparrow\downarrow,0 \rangle
\nonumber \\ U|\downarrow\uparrow,0
\rangle=&&|\downarrow\uparrow,0 \rangle \nonumber \\ \hat U
|\uparrow \uparrow,0 \rangle = && |\uparrow \uparrow,0  \rangle
\,\, ,
\end{eqnarray}
with $\Omega_{eff}=i \eta \Omega_{o} g_{o}^{2}(0) f_{1}(0) \lbrack
f_{o}(0) - f_{o}(1) \rbrack$. The QPG is implemented by letting
this evolution operator act, during a time interval equivalent to
a $2 \pi$ pulse, over the initial state
\begin{equation}
|\Phi\rangle = [\alpha|\downarrow\downarrow\rangle +
\beta|\downarrow\uparrow\rangle + \gamma|\uparrow\downarrow\rangle
+\delta|\uparrow\uparrow\rangle ] |0 \rangle ,
\end{equation}
obtaining
\begin{equation}
|\Phi^{\prime \prime} \rangle=[-\alpha|\downarrow\downarrow\rangle
+ \beta|\downarrow\uparrow\rangle
+\gamma|\uparrow\downarrow\rangle +\delta|\uparrow\uparrow\rangle
] |0 \rangle ,
\end{equation}
as desired. The CNOT gate could also be implemented, by means of
additional local operations, following the recipe given at the
beginning of this work.

The main feature of this method is that we were able to implement
a universal quantum logic gate with a single collective Rabi flip.
The interaction needs to be turned on just once with fixed
parameters. In fact, although two ions must be adressed individually
at the
same time, one needs just one laser beam split in two separated
pairs. These facts may be considered as advantages when comparing
our proposal to other ones requiring several consecutive and differently
adjusted interactions~\cite{ciraczoller1,sorensen,daniel}.
Although we are dealing with a dispersive interaction,
instead of a resonant one, which results in a slower process, we
are less exposed to errors arising from the sequential switching
of lasers. It is also worth noticing that the experimental tools
to implement many of these schemes, including the one proposed
here, are avaiable. An experiment from the NIST group produced 4
ions in a linear array, cooled down to their collective ground
state ~\cite{sacmar00}, and individual ionic addressing has been
achieved by the Innsbruck group~\cite{blatt}.

In the domain of CQED, we can find another possibility of making
use of a selective interaction for implementing a QPG. CQED has
already been recognized as a system where quantum logical
operations can be implemented~\cite{qpgparis,qpgkimble}. There are a
number of proposals where logical gates are performed either in
the electronic states of the atoms crossing the cavity or in a
combined system atom-field \cite{realizable}. We will be dealing
here with a system where logical operations are performed in
quantized modes of the cavity field. The two qubits are encoded in
two non degenerated modes of the electromagnetic field inside a
high $Q$ cavity that can have either one or zero photons. In this
case, atoms will serve only as catalizers of the logical
operation. Modes of the electromagnetic field in a cavity have
been proposed as a possible environment where quantum
logical operations can be done~\cite{tombesi}. Nevertheless, the
setup we will discuss here is different in some fundamental
aspects. While in \cite{tombesi} each qubit corresponds to one
mode inside a different cavity, in our scheme there are two non
degenerated modes inside a single cavity. Our scheme presents an
additional advantage, it is not necessary to make the atom cross
the cavity more than once. We will show that, again, a QPG can
be implemented with only a single interaction pulse of a three
level atom crossing the cavity.

The experimental setup considered is easily identified as the one
corresponding to a non degenerated two-photon
micromaser~\cite{micro}. The cavity is crossed by a three level
atom that interacts with the cavity field during a time interval
much smaller than the atomic and field decay time. Here, an
effective Hamiltonian corresponding to a two photon transition can
be derived in the limit where transitions from the lower
level $|g \rangle$ and upper level $|e \rangle$ to an intermediate level
$|i \rangle$ are not resonant, the detuning being such that
$\Delta=\omega_I-(E_e-E_i)/\hbar \gg
\frac{\Omega_{ei}^2+\Omega_{ig}^2}{\Delta}$ (see Fig. 2).
In this formula, $E_e$ and $E_i$ are the energies
of the levels $|e \rangle$ and $|i \rangle$, respectively and
$\Omega_{ei}$ ($\Omega_{ig}$) is the Rabi frequency corresponding
to the transition $|e\rangle \rightarrow |i\rangle$ ($|i \rangle
\rightarrow |g \rangle$). All these calculations are found in
\cite{micro}. \vspace*{0.5cm}
\begin{figure}
\leavevmode \epsfxsize=5 cm \epsfbox{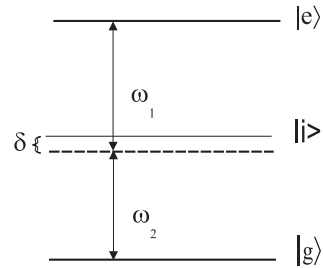}
\vspace*{0.5cm}
\caption{Three level atom interacting dispersively with two cavity
modes of different frequencies $\omega_1$ and $\omega_2$.}
\label{fig2}
\end{figure}
\noindent

Explicitly, the effective two-photon interaction
Hamiltonian reads, not writting the self energy terms,
\begin{equation}\label{hamcav}
\hat H_{\rm int}=\hbar \Omega \left (|e\rangle\langle g |\hat
a_1\hat a_2 + {\rm H.c.} \right),
\end{equation}
where $\Omega= \frac{\Omega_{ei}\Omega_{ig}}{\Delta}$ is the
effective Rabi frequency, $\Omega_{ei},\Omega_{ig}$ taken as real.
Supposing that each cavity mode has either one or zero photons and
that the decoupled atom is in the ground state, the most general
initial pure state for the combined atom-field system is
\begin{equation}\label{estcav}
|\psi \rangle = \left [\alpha |0,0\rangle
+\beta|1,0\rangle+\gamma|0,1\rangle+\delta|1,1\rangle \right
]|g\rangle.
\end{equation}
It is clear  that states in the subspace $\left \{|0,1\rangle,
|1,0\rangle \right \} |g \rangle$ and the ground state $|0,0\rangle
|g\rangle$ will not evolve under the interaction described in
Eq.~(\ref{hamcav}). Only $|1,1\rangle|g\rangle$ will suffer Rabi
oscillations with the effective frequency $\Omega$
\begin{equation}\label{evo}
|1,1\rangle |g \rangle \rightarrow \cos(\Omega t) |1,1\rangle
|g\rangle + \sin(\Omega t)|0,0\rangle|e\rangle.
\end{equation}
By selecting the atom velocity, in such a way that the interaction
time corresponds to a $\pi$ pulse, the final state
\begin{equation}\label{qpgcav}
|\psi^{\prime \prime} \rangle = \left [\alpha |0,0\rangle
+\beta|1,0\rangle+\gamma|0,1\rangle-\delta|1,1\rangle \right
]|g\rangle
\end{equation}
is produced after the atom leaves the cavity. This corresponds to
the action of a QPG on the state of Eq.~(\ref{estcav}). This
result shows an alternative way of implementing universal quantum
gates in the modes of the electromagnetic field in a high-$Q$ cavity
with a rather known scheme.

Both schemes presented here show an interesting feature about
quantum gates. Although they are proposed as logic gates for the
electronic leves of the trapped ions and the electomagnetic
modes of a cavity, they can be thought, as well, as
three-qubit quantum gates where the auxiliary CM vibronic
mode and the atomic electronic states, respectively,
are now the target qubits. In
this sense, these schemes perform a Deutsch gate
(controlled-controlled-rotation)~\cite{Deutsch} where the two qubits of
the ionic electronic levels and of the electromagnetic modes of the cavity
plays the rol of the control qubits. This
is a characteristic of logic gates implemented by quantum systems
where, since the operations are unitary, the process can be
regarded as two-way logic gates.

To summarize, we have shown that a selective interaction can be
useful for implementing quantum logic schemes in trapped ions and
in the domain of CQED. Specifically, we made a proposal for
implementing a QPG with a single pulse of a selective interaction
in these two systems. In both cases, we find that our scheme reduces
significantly the number of steps required for the gate operation,
which may be attractive when thinking in scalable quantum logical
processes.

E. Solano would like to thank N. Zagury and R. L. de
Matos Filho for useful comments about the ionic quantum gates. The
authors also acknowledge the support of Conselho Nacional de
Desenvolvimento Cient\'\i fico e Tecnol\'ogico (CNPq), Funda\c
c\~ao de Amparo \`a Pesquisa do Estado do Rio de Janeiro (FAPERJ),
Funda\c c\~ao Universit\'aria Jos\'e Bonif\'acio (FUJB), and
Programa de Apoio a N\'ucleos de Excel\^encia (PRONEX).

$^*$ Present address: Max-Planck Institute for Quantum Optics,
Hans-Kopfermann Strasse 1, D-85748 Garching, Germany (Enrique.Solano@
mpq.mpg.de).

$^{**}$ Present address: Laboratoire Kastler Brossel, D{\'e}partement
de Physique de l'Ecole Normale Sup{\'e}rieure, 24 rue Lhomond, F-75231
Paris Cedex 05, France.

\end{document}